\documentclass[letterpaper, 10 pt, journal, twoside]{IEEEtran}

\IEEEoverridecommandlockouts                              % This command is only needed if 

\usepackage{xcolor}
\usepackage{graphics} % for pdf, bitmapped graphics files
\usepackage{epsfig} % for postscript graphics files
\usepackage{mathptmx} % assumes new font selection scheme installed
\usepackage{times} % assumes new font selection scheme installed
\usepackage{amsmath} % assumes amsmath package installed
\usepackage{amssymb}  % assumes amsmath package installed
\usepackage[switch]{lineno}
\usepackage{multirow}
\usepackage{booktabs}
\usepackage{acro}
\usepackage{amssymb}
\usepackage{bm}
\usepackage{algorithm}
\usepackage{algorithmicx}
\usepackage{algpseudocode}
\usepackage{soul}
\usepackage{hyperref} 
\usepackage{esvect}
\usepackage{stfloats}
\usepackage{nicefrac}
\usepackage{url}
\usepackage{cite}

\usepackage{titlesec}
\titlespacing{\section}{0pt}{*2.5}{*0.5}
\titlespacing{\subsection}{0pt}{*1.5}{*0.5}
\begin{document}

\title{Dexterous Intramyocardial Needle Ablation (d-INA): Design, Fabrication, and \\In-Vivo Validation}

\author{Chang Zhou$^{1*}$, Charles P. Hong$^{2*}$, Yifan Wang$^{1}$, Ehud~J.~Schmidt$^{3}$, Junichi~Tokuda$^{4}$,  Aravindan~Kolandaivelu$^{3\dagger}$ and Yue Chen$^{2\dagger}$,~\IEEEmembership{Member,~IEEE}% <-this % stops a space
\thanks{This work was supported by the NSF CAREER Award 2339202, and the National Institute of Biomedical Imaging And Bioengineering of the National Institutes of Health (NIH) under Award Number R01EB034359.  This work involved  animals in its research. Approval of all ethical and experimental procedures and protocols was granted by JHU’s Institutional Animal Care and Use Committee (IACUC) with the protocol number of SW23M128.}

\thanks{$^{1}$C. Zhou and Y. Wang are with the Department of Mechanical Engineering, 
        Georgia Institute of Technology, Atlanta, GA 30332 USA ({\href{mailto:czhou368@gatech.edu}{czhou368@gatech.edu}} and {\href{mailto:wangyf@gatech.edu}{wangyf@gatech.edu}}). 
        }%
        
\thanks{$^{2}$C. P. Hong and Y. Chen are with the Department of Biomedical Engineering, Georgia Institute of Technology $/$ Emory University, Atlanta  GA 30332 USA ({\href{mailto:chong71@gatech.edu}{chong71@gatech.edu}})and {\href{mailto:yue.chen@bme.gatech.edu}{yue.chen@bme.gatech.edu}}.}%

\thanks{$^{3}$E. J. Schmidt and A. Kolandaivelu are with the Department
of Medicine, Johns Hopkins
University, Baltimore MD 21205 USA (e-mail: {{\href{mailto:eschmi17@jhu.edu}{eschmi17@jhu.edu}} and \href{mailto:akoland1@jhmi.edu}{akoland1@jhmi.edu}}).}% <-this % stops a space

\thanks{$^4$J. Tokuda is with the Department of Radiology, Harvard Medical
School, Boston MA 02115 USA. ({\href{mailto:tokuda@bwh.harvard.edu}{tokuda@bwh.harvard.edu}}).}

\thanks{$^*$C. Zhou and C. P. Hong contributed equally to this work.}
\thanks{$^\dagger${Corresponding author: Aravindan~Kolandaivelu and Yue~Chen.}}

 }

\setlength{\textfloatsep}{1.pt}% Remove \textfloatsep

% % The paper headers
% \markboth{Journal of \LaTeX\ Class Files,~Vol.~14, No.~8, August~2015}%
% {Shell \MakeLowercase{\textit{et al.}}: Bare Demo of IEEEtran.cls for IEEE Journals}
%\markboth{IEEE Robotics and Automation Letters. Preprint Version. Accepted December, 2024}
%{Cai \MakeLowercase{\textit{et al.}}: Modular Self-Reconfigurable Continuum Robot for General Purpose Loco-Manipulation} 

\maketitle

%%%%%%%%%%%%%%%%%%%%%%%%%%%%%%%%%%%%%%%%%%%%%%%%%%%%%%%%%%%%%%%%%%%%%%%%%%%%%%%%
\begin{abstract}
Radiofrequency ablation is widely used to prevent ventricular tachycardia (VT) by creating lesions to inhibit arrhythmias; however, the current surface ablation catheters are limited in creating lesions that are deeper within the left ventricle (LV) wall. 
%Intramyocardial needle ablation (INA) uses a needle that penetrates the ventricular wall and delivers ablation from within. 
Intramyocardial needle ablation (INA) addresses this limitation by penetrating the myocardium and delivering energy from within. 
Yet, existing INA catheters lack adequate dexterity to navigate the highly asymmetric, trabeculated LV chamber and steer around papillary structures, limiting precise  targeting.
%Yet, existing INA catheters' sub-optimal dexterity limits their capability to navigate within the asymmetric ventricular chamber with trabeculations and papillary muscles. 
This work presents a novel dexterous INA (d-INA) toolset designed to enable effective manipulation and creation of deep ablation lesions. The system consists of an outer sheath and an inner catheter, both bidirectionally steerable, along with an integrated ablation needle assembly. Benchtop tests demonstrated that the sheath and catheter reached maximum bending curvatures of 0.088 mm$^{-1}$ and 0.114 mm$^{-1}$, respectively, and achieved stable ``C", ``S", and non-planar ``S" shaped configurations. 
Ex-vivo studies validated the system’s  stiffness modulation  and lesion creation capabilities. 
In-vivo experiments in two swine demonstrated the device’s ability to reach previously hard-to-access locations such as the LV summit region, and achieved a 219\% increase in ablation depth compared with a standard ablation catheter. 
These results establish the proposed d-INA as a promising platform for achieving deep ablation with enhanced dexterity, advancing the treatment for VT.

\end{abstract}


% \IEEEpeerreviewmaketitle

%%%%%%%%%%%%%%%%%%%%%%%%%%%%%%%%%%%%%%%%%%%%%%%%%%%%%%%%%%%%%%%%%%%%%%%%%%%%%%%%

\section{Introduction}

\IEEEPARstart{V}{entricular} tachycardia (VT) is a rapid, potentially lethal heart rhythm and a major cause of sudden cardiac death (SCD) which kills 325,000 people (US) per year \cite{jaramillo2023sudden,kurl2021exercise,martin20252025}. It commonly arises in the context of ventricular scar, often following myocardial infarction \cite{shivkumar2019catheter, hadjis2025ventricular}. The most effective therapy for recurrent VT is electro-physiological catheter ablation, which typically uses radiofrequency (RF) energy to thermally induce necrosis of the aberrantly conducting tissue that leads to VT. The goal of ablative treatment is to create localized permanent tissue injury in aberrant electrical foci or propagation pathways, causing arrhythmia. Ablation energy is applied to accessible parts of the endocardial or epicardial surface near the slowest propagating part of the VT conduction pathway. A series of ablation lesions attempts to transect this isthmus to prevent abnormal propagation. Multiple randomized controlled trials and meta-analyses demonstrate the potential of RF ablation for VT treatment. However, catheter VT ablation is currently only about 60\% successful at one year \cite{atreya2022best,elbatran2021contact,subramanian2023catheter,sapp2016ventricular,dinov2014outcomes}.

Multiple factors contributed to these sub-optimal treatments. First, the conventional RF catheter is unable to ablate deeper within the myocardium. Specifically, ablation typically applies RF current through an ablation catheter tip contacting the myocardial wall surface, causing thermally induced necrosis above 50$^\circ\mathrm{C}$ \cite{nath1993cellular}. Heat conduction away from the catheter tip limits ablation lesion size, especially through scar and fat, which have lower thermal conductivity \cite{grayson1967thermal, liu2006characterization}. Second, it is challenging to systematically and efficiently ablate multiple VT propagation isthmuses in the dynamic ventricular chamber. Individuals with VT often have multiple aberrant pathways, and not ablating all pathways reduces treatment success \cite{santangeli2014end}. Improving procedure success by ablating more tissue is limited by long procedure durations of 4-8 hours \cite{yu2015catheter}. Individuals with VT also typically have reduced cardiac contractile function, increasing the risk of hemodynamic instability and complications with longer procedures \cite{yu2015catheter,turagam2017hemodynamic}.

Intramyocardial needle ablation (INA) is an effective intramyocardial ablation method that can deliver ablation energy towards the deeply seated targets \cite{sapp2013initial,sapp2006large,packer2022ablation}. However, INA is not routinely used as a first-line strategy in the initial VT ablation procedures due to the cumbersome process required to systematically move the INA catheter across the non-smooth left ventricle (LV) surface, hindering the ablation of multiple VT target sites. The INA device, originally intended for intramyocardial (IM) drug delivery, features a straight needle that extends from the steerable catheter tip and is inserted into the LV wall for thermal delivery \cite{stevenson2019infusion,schaeffer2020characteristics}. Despite successful validations in the multi-center human trials, the current toolset remains suboptimal for INA ablation treatment. One of the primary limitations is the compromised dexterity since it is challenging for the clinician to insert the device at the desired location and at the same time maintain perpendicular to the wall surface, which is necessary for effective lesion formation. The limited torquability and steerability of the INA catheter further restrict access to anatomically challenging regions such as the papillary muscles and the basal LV summit \cite{mariani2022catheter,nakajima2021periaortic,ouyang2014ventricular}.

To address the above-mentioned limitations, we developed a novel dexterous INA (d-INA)-based VT treatment system prototype with the following features: i) dexterous motion that can exhibit both planar and non-planar “S” shape to access the hard-to-reach locations with the desired tip orientation, ii) enhanced torquability to enable precise and effective axial control in the tortuous and constrained cardiac chambers. 
This work aims to address the limitations of conventional RF ablation catheters and existing INA catheters by introducing a design that can achieve superior dexterity and mechanical controllability. 
%This work holds the potential  addressing the limitations of conventional RF ablation by creating deep ablations and current INA catheters by introducing a design that can achieve superior dexterity and mechanical controllability. 
This paper is arranged as follows. Section \ref{sec:MaterialsandMethods} presents the working principle of the proposed d-INA design and fabrication method. Section \ref{sec:BenchTopExperimentalValidations} describes the benchtop characterization results, followed by in-vivo experimental validations in Section \ref{sec:InVivoExperimentalValidations}. Finally, the conclusions and future work are presented in Section \ref{sec:conclusion}.

% followed by discussion and conclusion in Section IV.  

% , and iii) MRI-guided lesion assessment to facilitate future intra-procedural MRI VT ablation. 

%\section{Related work}
%\input{section/2.related.tex}

\section{Materials and Methods}
\label{sec:MaterialsandMethods}
This section highlights the working principle, design specifications, and manufacturing process for the proposed d-INA system. 

\subsection{d-INA Design Overview}

\begin{figure*}[!t]
\centering
\includegraphics[width=0.8\linewidth]{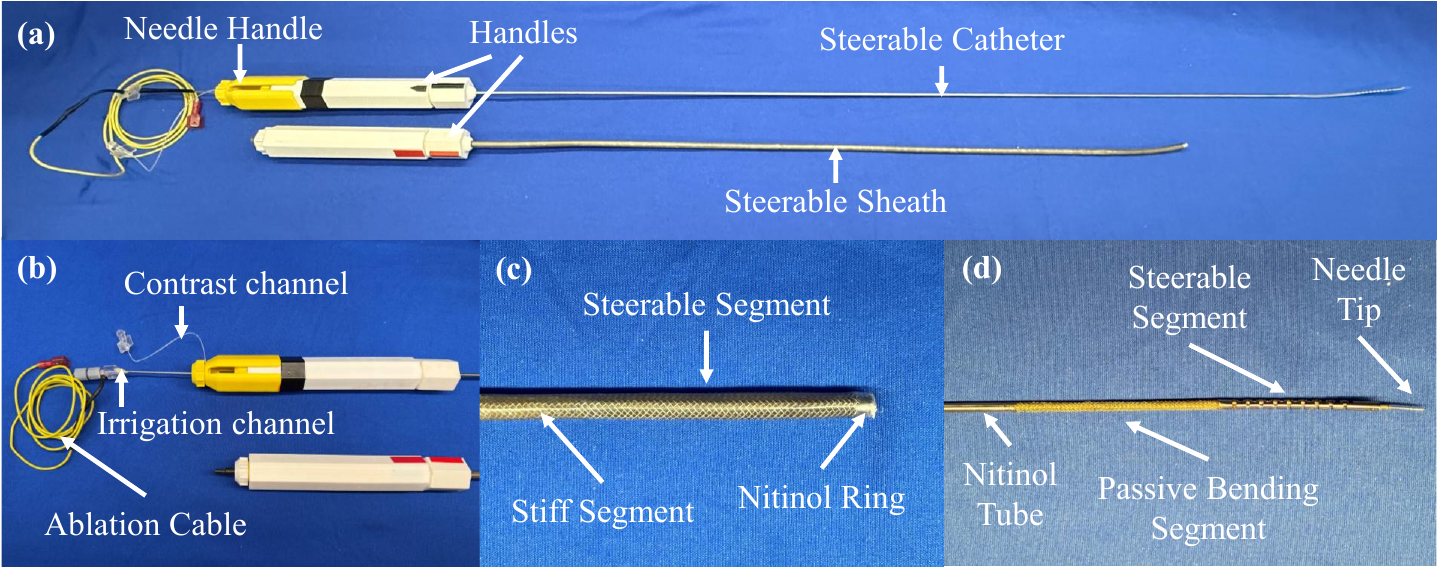}
\caption{(a) The proposed d-INA toolset with a steerable outer sheath and an integrated catheter/needle system. (b) Detailed view of the proximal tip of the d-INA toolset. (c) Detailed view of the distal tip of the outer sheath. (d) Detailed view of the distal tip of the inner catheter.}
\label{fig:prototype}
\end{figure*}%\vspace{-2pt}

Figure \ref{fig:prototype} shows the prototype of the proposed d-INA toolset, which consists of an outer sheath, an inner catheter, and an ablation needle that can penetrate the tissue for deep ablation. Both the outer sheath and the inner catheter possess three degrees of freedom (DoF) to enable linear insertion, axial rotation, and tip deflection. The ablation needle is meant to be inserted from the inner catheter towards intramyocardial VT targets for RF energy delivery.

\begin{figure}[!t]
\centering
\includegraphics[width=\linewidth]{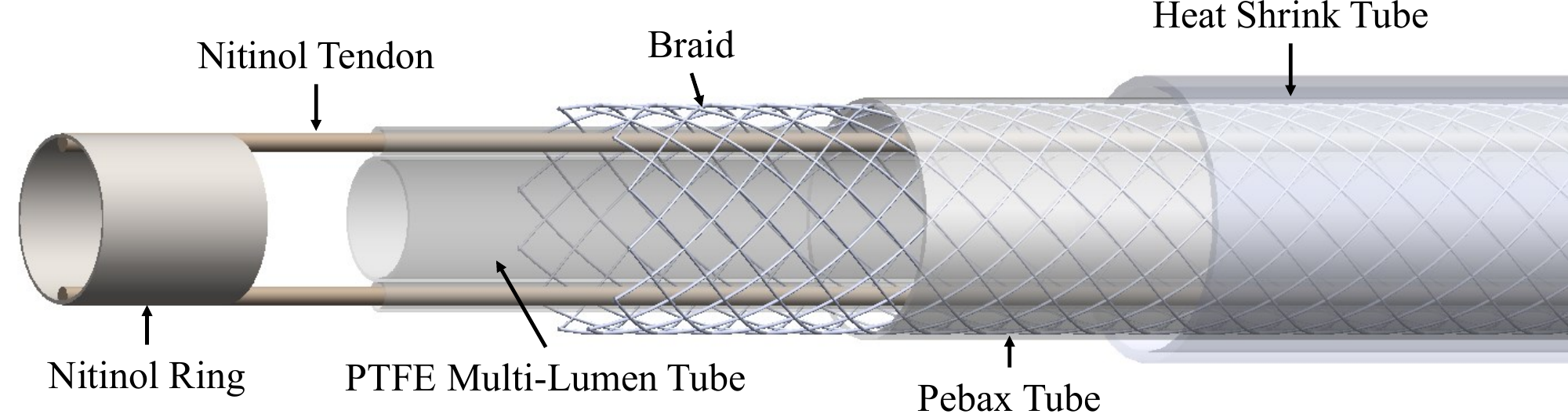}
\caption{The detailed design of the steerable outer sheath.}
\label{fig:sheath}
\end{figure}

\subsection{Steerable Outer Sheath Design} 
\vspace{-2pt}

The steerable sheath houses and provides the guidance for the inner catheter/needle manipulation within the LV. It has an outer diameter (OD) of 5 mm, an inner diameter (ID) of 3 mm for passing the inner tools, and a total length of 700 mm. The detailed structure of the outer sheath is illustrated in Figure \ref{fig:sheath}. The inner-most layer is a polytetrafluoroethylene (PTFE) multi-lumen liner (SLW MultiLumen Etch, ZEUS, USA), forming the structural core of the sheath. The PTFE liner has a 3 mm ID main lumen for passing the inner catheter and two 0.6 mm ID satellite lumens distributed on opposite sides of the main lumen for passing two 0.5 mm OD nitinol tendon wires. The liner is sheathed in a tinned copper braid (Braid 2162, Alpha Wire, USA) to enhance torsional rigidity. Two Pebax tubes with different durometers (i.e., 35D and 72D) are melted onto the multi-lumen liner via a reflow process to bind it with the copper braid, achieving different stiffness along the sheath. Specifically, we aim to ensure the distal tip has a relatively lower bending stiffness such that the bending motion is concentrated near the tip with tendon input, achieving higher curvature and thus better steerability. To achieve bi-directional bending, two tendons are laser-welded to a nitinol ring with 4 mm OD and 3.8 mm ID, which is then attached to the distal tip of the multi-lumen liner. The stiff segment has a length of 650 mm, while the steerable segment has a length of 50 mm. During the reflow process, a heat shrink tube is used outside of the Pebax tubes to simultaneously ensure a smooth surface finish and control the final OD of the steerable sheath.

\subsection{Steerable Inner Catheter Design}

\begin{figure*}[!t]
\centering
\includegraphics[width=0.85\linewidth]{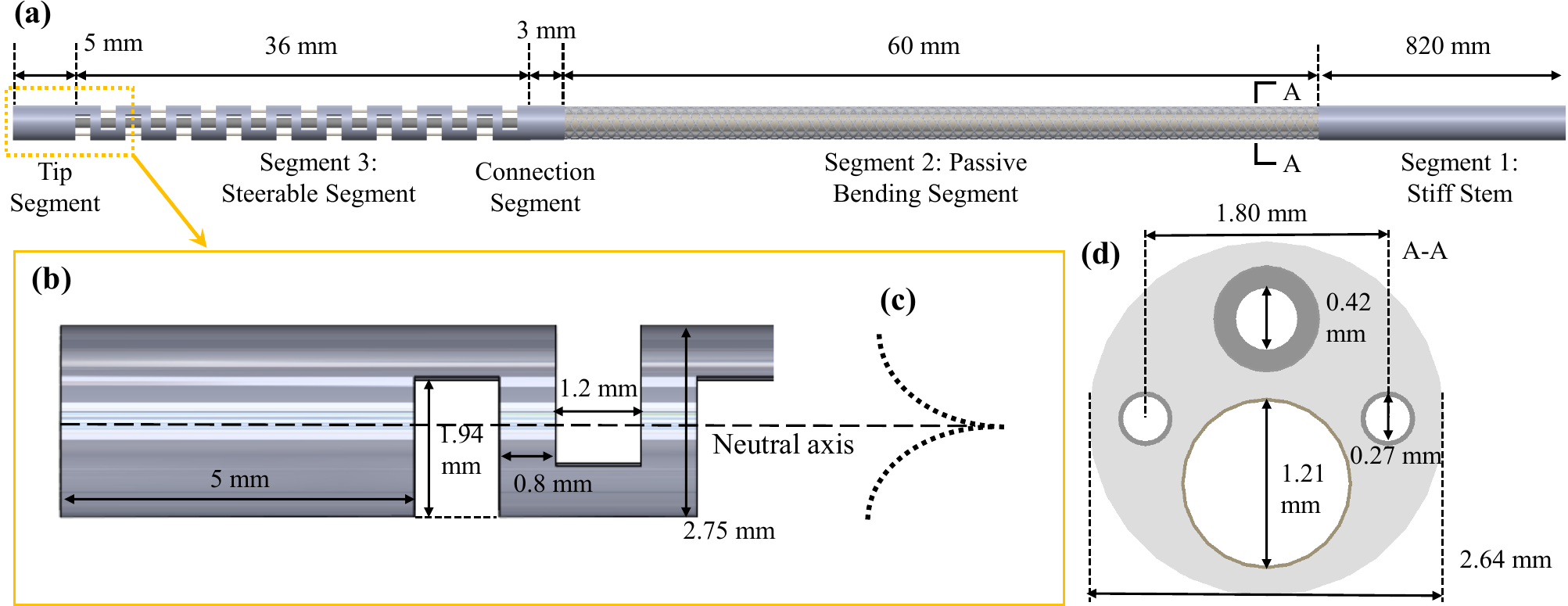}
\caption{The CAD design of the steerable catheter. (a) The detailed view of the steerable catheter distal tip. (b) Zoomed-in view of the notched Nitinol tube with parameters. (c) Schematic diagram showing the steerable catheter bending directions. (d) The cross-section of the passive bending segment.}
\label{fig:catheter}
\end{figure*}

The steerable catheter is designed to provide a larger workspace, enable dexterous manipulation, and serve as an insertion channel for the needle. The detailed design of the catheter distal tip is shown in Figure \ref{fig:catheter}(a). The catheter consists of three segments: (i) a stiff stem, (ii) a passive bending segment, and (iii) a steerable segment. The stiff stem is made of a Nitinol tube to provide high torsional and bending stiffness. The passive bending segment is a composite structure combining a multi-lumen tube and metallic braid reinforcement for intermediate torsional and bending stiffness. The steerable segment is a notched Nitinol tube that achieves low bending stiffness for large bending curvature. Again, such a design concentrates the bending motion near the tip, achieving better distal steerability. Specifically, the steerable segment is a bi-directional asymmetric notched Nitinol tube with a 36 mm bendable length and 0.1 mm wall thickness, precisely patterned using a femtosecond laser. The parameters of the notches are shown in Figure \ref{fig:catheter}(a). On the distal end of the notched Nitinol tube, two Nitinol tendons of 0.25 mm OD are laser-welded onto the 5 mm uncut segment at the tip. On the proximal end, a 3 mm uncut segment connects the bending segment with the passive bending segment via reflowing. 

While the notched design facilitates high planar bending curvature, it lacks the capability to bend outside the notch plane (see Figure \ref{fig:catheter}(c) for the steerable catheter bending directions). To achieve non-planar bending when the sheath and catheter are assembled, a passive bending segment is included. Specifically, the outer sheath establishes the first planar bending direction during operation, while the passive bending segment of the catheter aligns with the steerable segment of the sheath and contorts to the established first bending direction. The steerable segment of the catheter is extended beyond the tip of the sheath and can rotate freely, establishing the second planar bending direction. The 60 mm passive bending segment is fabricated based on a Pebax 55D multi-lumen tube with four lumens, as shown in Figure \ref{fig:catheter}(d). The dimensions and functions of each lumen are as follows: a 1.21 mm ID lumen for the ablation needle, a 0.42 mm ID lumen for contrast agent injection, and two 0.27 mm ID lumens for the Nitinol tendons. To enhance torsional stiffness, the multi-lumen tube is sheathed in a tinned copper braid reinforcement (Braid 1223, Alpha Wire, USA). A PTFE liner is positioned within each lumen to reduce the friction between the inner lumen wall and the sliding components inside, such as tendons and an ablation needle. Similar to the steerable sheath fabrication, the multi-lumen tube, braid, and PTFE liners are bound together through the reflow process.

\subsection{Ablation Needle Development}

\begin{figure}[!t]
\centering
\includegraphics[width=\linewidth]{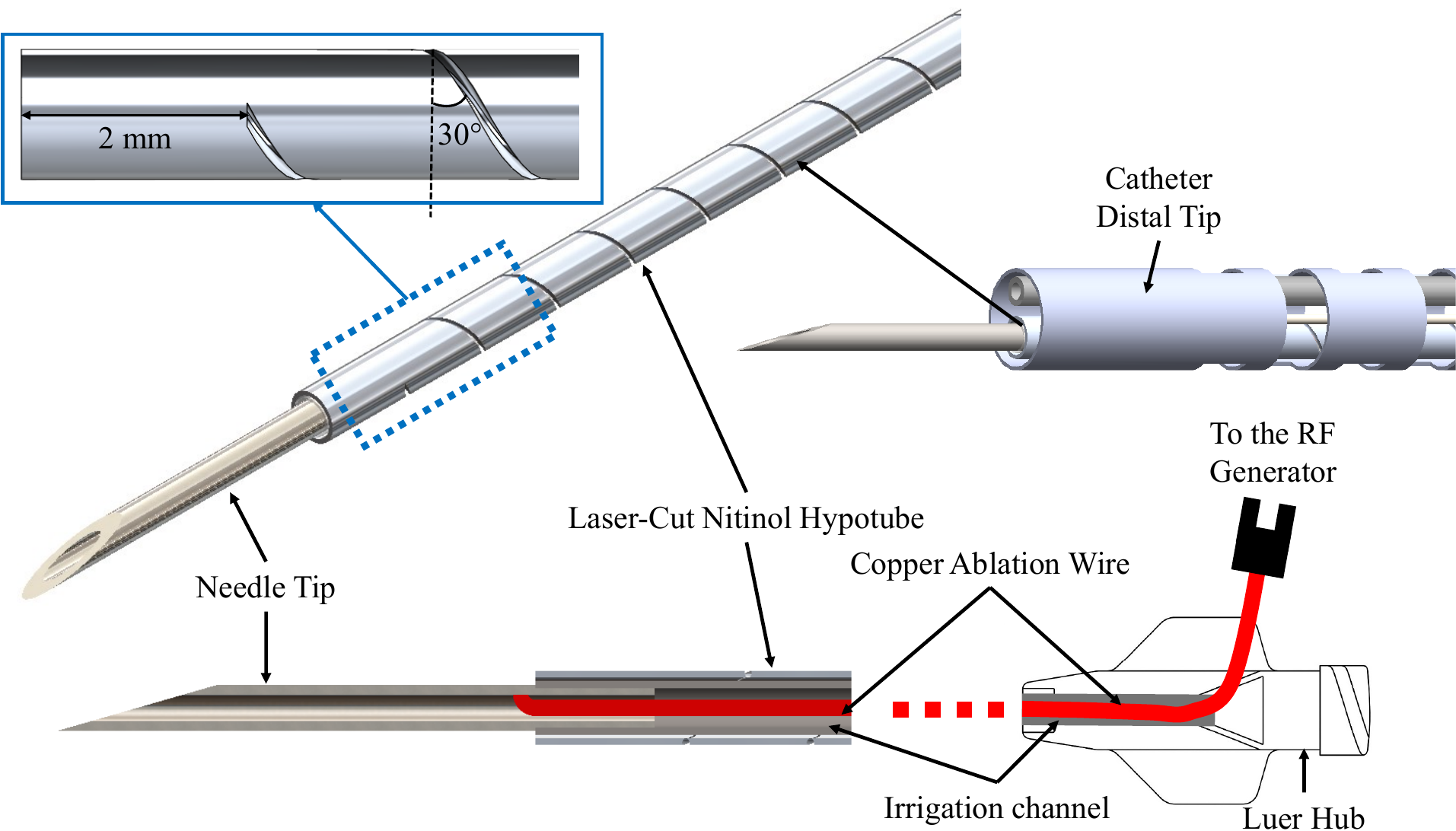}
\caption{Schematic of the ablation needle sub-system.}
\label{fig:needle}
\end{figure}

The ablation needle sub-system is comprised of four components: a standard 22 AWG hypodermic needle, a 32 AWG insulated copper wire, an irrigation tube, and a laser-cut Nitinol hypotube. The ablation needle assembly is showcased in Figure \ref{fig:needle}. 

The hypodermic needle is cut to an 8 mm segment, with a 3 mm portion welded to the hypotube to facilitate penetration. The copper wire is threaded through the Rilsamid AMNO irrigation tube with 0.62 mm ID and 0.85 mm OD, into the end of the hypodermic needle, and laser-welded directly to the needle to pass the RF current from the generator to the needle tip. The irrigation tube and the hypodermic needle are affixed using epoxy. The final construct is secured within the laser-cut Nitinol hypotube with 0.62 mm ID and 0.85 mm OD, which serves as the structural backbone. At the base of the hypotube, a modified Luer hub is attached to the assembly. Specifically, a bore is added to the top of the hub, allowing the passage of the ablation wire, which is then directly connected to an RF generator. A syringe can be attached to the Luer hub to provide irrigation to the tip of the needle to enable large lesion creation. Epoxy is applied to all connection points to prevent leakage. Once assembled, the ablation needle is advanced into the inner catheter until its tip aligns flush with the distal catheter end, which encapsulates the needle tip to minimize perforation risk.

Functionally, the hypotube is laser-cut with a spiral pattern ($30^\circ$ helix angle) to achieve sufficient compliance for passive bending within a steerable sheath and catheter. This inherent flexibility ensures that the needle assembly does not compromise the system's steerability during manipulation. Once the sheath and catheter are positioned at the desired location, the needle tip is pushed from the back and deployed from the catheter to penetrate the cardiac tissue, delivering ablation energy directly to deep targets with the integrated irrigation channel providing continuous flow.

\subsection{Steerable Handle Design }

\begin{figure}[!t]
\centering
\includegraphics[width=\linewidth]{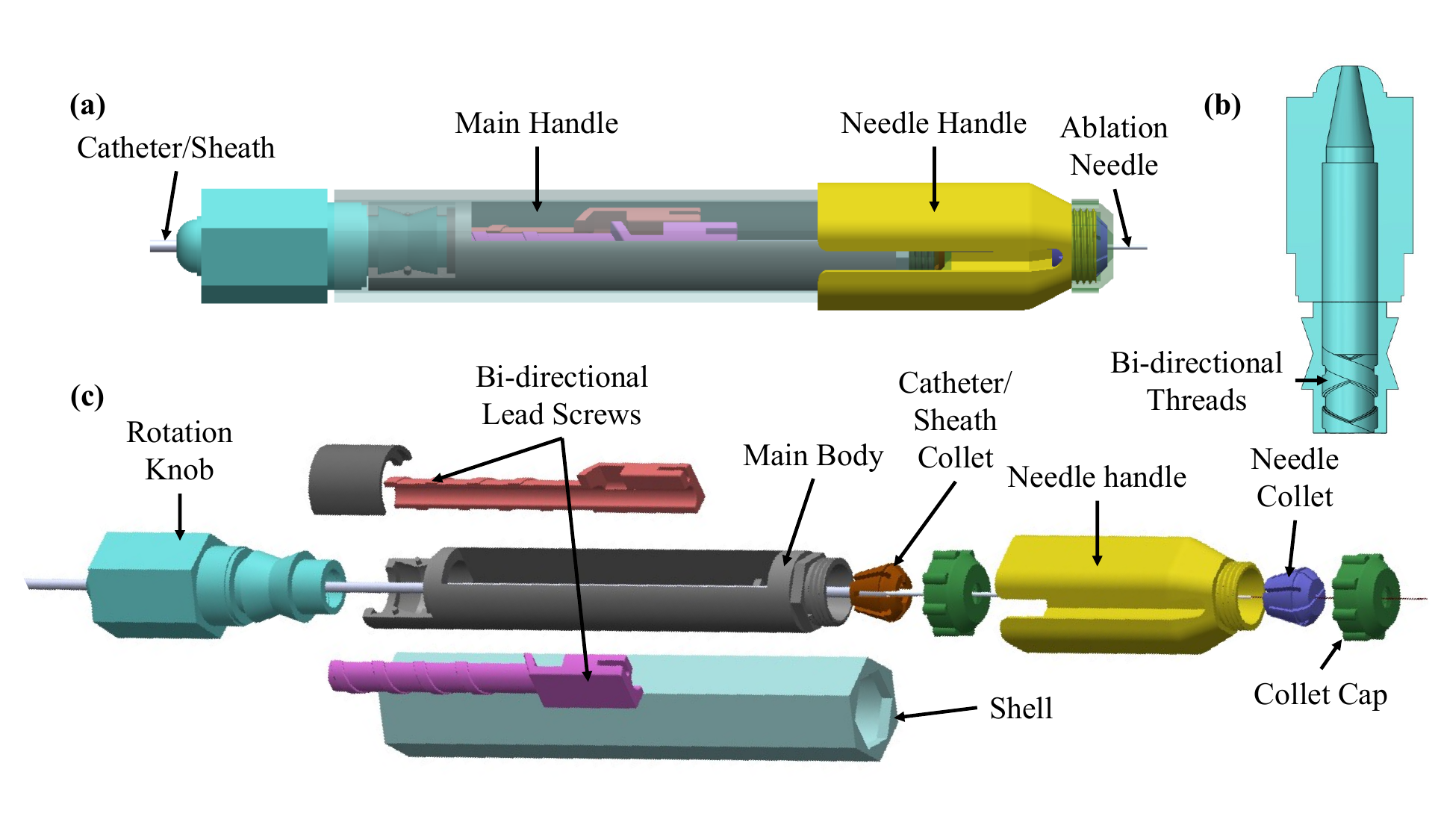}
\caption{CAD model of the handle design. (a) The handle assembly. (b) The section view of the rotation knob. (c) The exploded view of the handle assembly.}
\label{fig:handle}
\end{figure}

The steerable catheter and sheath are actuated by an identical handle mechanism, illustrated in Figure \ref{fig:handle}. The handle incorporates a bi-directional lead screw mechanism for tendon control, a rotation knob to actuate the lead screws, a hexagonal shell housing the assembly, and a collet mechanism for fixing the catheter or sheath. The core component of the design is the bi-directional lead screw, which comprises two half-cylinder lead screws with opposing threads: one right-handed and one left-handed. These screws engage with corresponding bi-directional internal threads within the rotation knob (Figure \ref{fig:handle}(b)). Tendon wires are attached to the extruded bases of the lead screws and terminated by aluminum crimp collars. Rotating the knob drives the two lead screws to translate in opposite directions with equal displacement, achieving a differential drive of the tendon pair to induce controlled bending in the steerable segments. % To ensure MRI compatibility, all handle components are 3D-printed from PLA.

%An additional needle handle is attached to the inner catheter handle, which interfaces with the hexagonal main handle body and can slide axially to control the insertion of the ablation needle. 
An auxiliary needle handle is integrated with the inner catheter handle, interfacing with the hexagonal shell and enabling axial sliding to precisely control ablation needle insertion. 
The needle handle is designed with a collet mechanism to securely lock the ablation needle. With this integrated design, the maximum needle insertion depth (i.e., 5 mm) is mechanically limited by the interaction between the hexagonal shell and the needle handle. This built-in mechanical stop helps prevent tissue perforation and enhances procedural safety.

\section{Bench-top Experimental Validations}
\label{sec:BenchTopExperimentalValidations}
A series of benchtop experiments were conducted to evaluate the capabilities of the developed d-INA system in terms of dexterous manipulation and ablation capability. %Upon successful benchtop verification, the toolset was tested in the animal trial. 

\subsection{Benchtop Validation} 

\begin{figure*}[!t]
\centering
\includegraphics[width=0.8\linewidth]{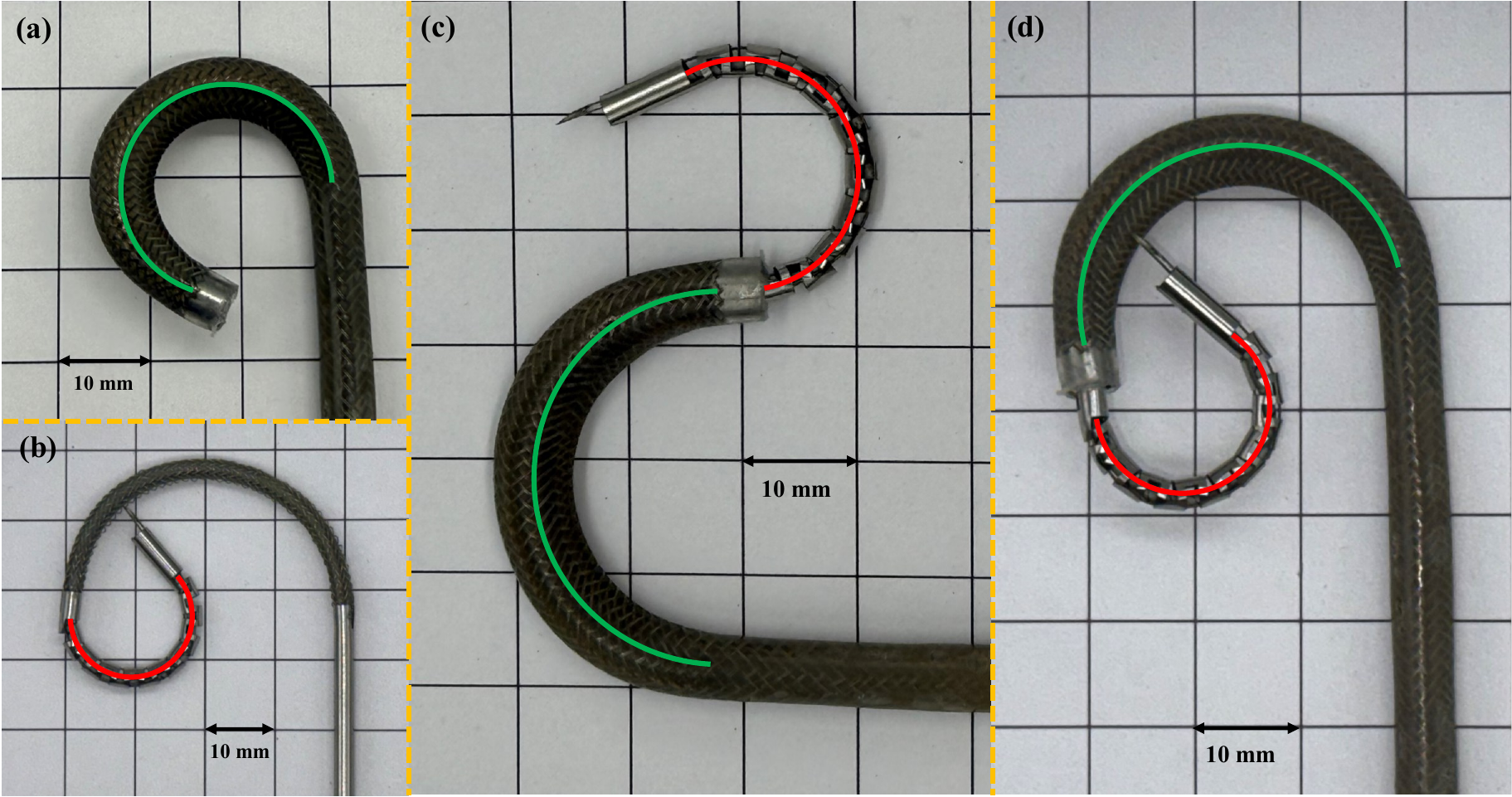}
\caption{The bending test results of the proposed d-INA device. The fitted centerlines of the sheath are shown in green, and the fitted centerlines of the catheter are shown in red. (a) The maximum bending of the sheath. (b) The maximum bending of the catheter. (c) The maximum "S" shape bending when the catheter is inside the sheath. (d) The maximum "C" shape bending when the catheter is inside the sheath.}
\label{fig:bending}
\end{figure*}

\begin{figure}[!t]
\centering
\includegraphics[width=\linewidth]{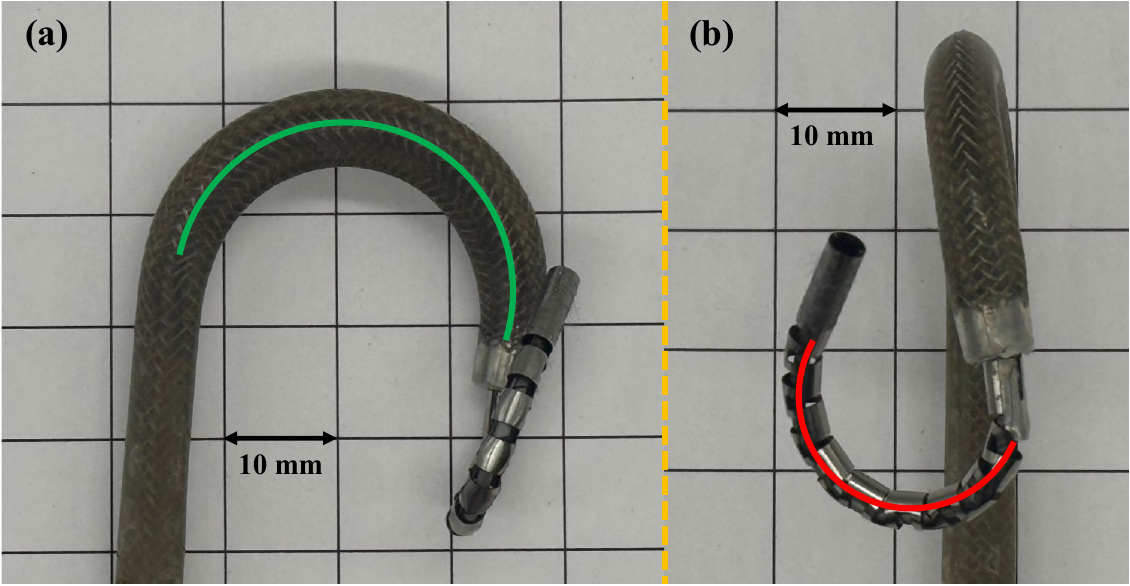}
\caption{Non-planar bending of the proposed d-INA toolset. (a) The front view of the non-planar bending shape. (b) The side view of the non-planar bending shape. }
\label{fig:ofp_bending}
\end{figure}

The steerability of the toolset was evaluated via bending curvatures of different segments, as shown in Figure \ref{fig:bending} and \ref{fig:ofp_bending}. Each tool was first evaluated separately, and then the combined assembly was evaluated to understand the effects of the combined stiffness. When the toolset is combined, we measured three extreme cases, namely, a planar ``S'' shape, a planar ``C'' shape, and a non-planar bending configuration with orthogonal bending planes. The minimum bending radii and maximum curvatures are obtained by fitting circular arcs to centerlines of the d-INA and reported in Table \ref{tab:curvature}. While the combination of the sheath and catheter has minimal effects on the maximum curvature of the steerable segment of the catheter, the maximum curvature of the sheath is decreased due to the increased overall bending stiffness. The worst-case scenario is the planar ``S'' shape configuration, where the sheath tendons need to compensate for the bending moment of the catheter in the opposite direction. Even in this worst case, the sheath achieved nearly $180^\circ$ bending angle, to enable entry into the LV from the LA, even in the small swine heart. Furthermore, owing to the passive bending segment incorporated in the inner catheter, the proposed d-INA is also able to achieve a stable non-planar bending configuration, as illustrated in Figure \ref{fig:ofp_bending}.  Steerability of the proposed d-INA is not affected by non-planar bending configurations. 

\begin{table*}[htbp]
\centering
\caption{The minimum bending radius and maximum bending curvature for the steerable segments of the outer sheath and inner catheter. }
\begin{tabular}{lcccc}
\hline
 & \multicolumn{2}{c}{\textbf{Outer Sheath}} & \multicolumn{2}{c}{\textbf{Inner Catheter}} \\
\cline{2-5}
\textbf{Configuration} & \textbf{Min. Radius (mm)} & \textbf{Max. Curvature (mm$^{-1}$)} & \textbf{Min. Radius (mm)} & \textbf{Max. Curvature (mm$^{-1}$)} \\
\hline
Individual Component & 11.33 & 0.088 & 8.75 & 0.114 \\
S Shape & 15.95 & 0.063 & 9.89 & 0.101 \\
C Shape & 15.11 & 0.066 & 8.14 & 0.123 \\
Orthogonal Bending Planes & 14.80 & 0.068 & 9.15 & 0.110 \\
\hline
\end{tabular}
\label{tab:curvature}
\end{table*}

\begin{figure*}[!t]
\centering
\includegraphics[width=0.95\linewidth]{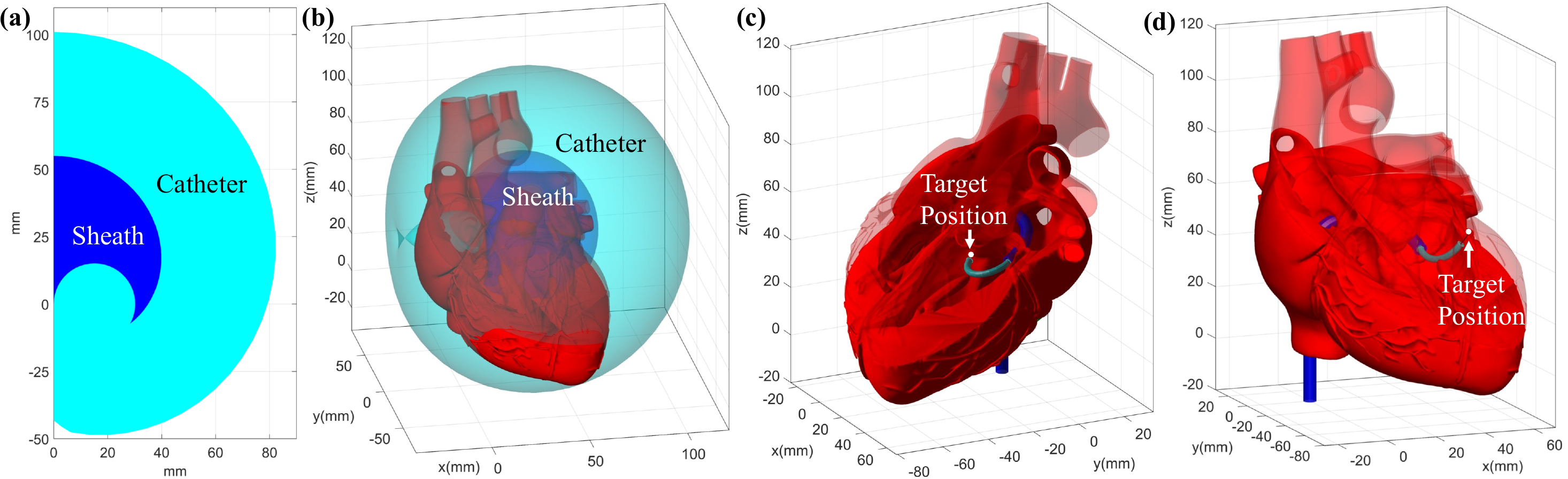}
\caption{The workspace analysis of the proposed d-INA based on the constant curvature assumption. (a) A 2D representation of the workspace of both the tip of the sheath (blue) and the tip of the ablation needle (cyan). The d-INA can freely rotate about the stem, thus generating a rotationally symmetric workspace. (b) The 3D workspace volume overlaid onto a human heart model. The upper half of the heart model is semi-transparent to show the internal chambers. (c) A typical ``C" shape configuration of the device reaching the basal septal and inferior region. The sheath is shown in blue, the catheter is shown in cyan, and the ablation needle is shown as a black line segment. (d) A typical ``S" shape configuration of the device reaching the LV outflow tract and basal lateral LV region.}
\label{fig:workspace}
\end{figure*}

Based on results from the bending test and the lengths of the steerable segments of the catheter and sheath, we generated the workspace of the d-INA, as shown in Figure \ref{fig:workspace}(a) and (b). The kinematics model is based on the constant curvature assumption \cite{webster2010constant_curvature}. The base of the workspace, corresponding to the origin in Figure \ref{fig:workspace}(a), is defined as the puncture point on the trans-atrial septal wall, with the penetration direction perpendicular to the sagittal plane. The device centerline is assumed to be tangent to the penetration direction at the base. It is also assumed that only the steerable segment of the sheath can be inserted through the puncture point. The workspace is generated by grid-sampling the planar configurations of the device within the length limits and the curvature limits obtained from the planar ``C'' shape bending test, obtaining the 2D envelope of all sampled positions (Figure \ref{fig:workspace}(a)), and rotating the envelope around the stem axis (Figure \ref{fig:workspace}(b)). 
The 3D workspace generated is further overlaid onto a human heart model, as shown in Figure \ref{fig:workspace}(b). The workspace of the sheath only covers the entire LA and some top regions of the LV, while the workspace of the catheter covers the entire LV. 

%We note that the papillary muscles and tendinous cords in the LV might constrain the workspace of the d-INA in practice. This could be partially overcome by leveraging flexible contacts between the catheter and the tissue, although this would render the constant curvature assumption less valid and require more sophisticated motion planning, as shown in our prior work \cite{wang2025topology}. Overall, the result suggests that the device has sufficient reach to perform INA throughout the LV chamber. 

As a further demonstration of the reachability of the device, two typical configurations of the device during the INA procedure are simulated within the heart model. The first configuration is a ``C" shape bending, as shown in Figure \ref{fig:workspace}(c). The sheath bends $72^\circ$ through the mitral valve, and the catheter bends $160^\circ$ in the same direction to reach the basal septal region. The second configuration is a ``S" shape bending, as shown in Figure \ref{fig:workspace}(d). The sheath bends $90^\circ$ through the mitral valve, and the catheter bends $90^\circ$ to reach the basal lateral LV region. The bending planes of the sheath and the catheter is $117^\circ$ apart for the "S" shape bending. 

\subsection{Ex-Vivo Multi-Configuration Penetration Experiment} 

\begin{figure*}[!t]
\centering
\includegraphics[width=\linewidth]{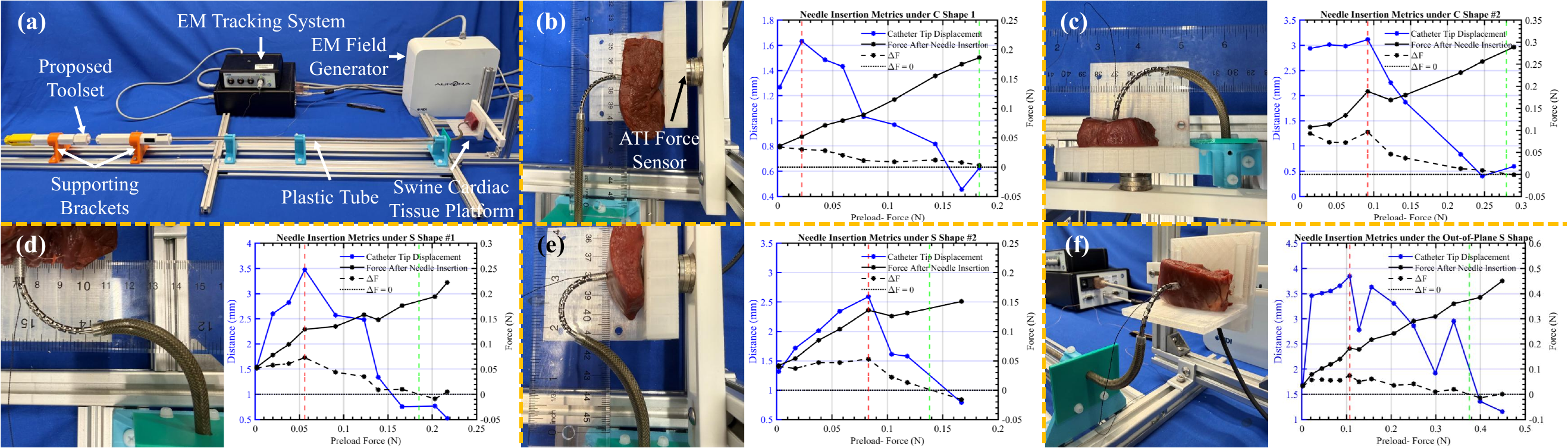}
\caption{Benchtop experimental setups and results for validating the needle puncture capability of the d-INA. (a) Experimental setup for measuring the required preload force for puncture, the force after needle insertion, and the catheter tip displacement. (b)-(c) Different bending configurations of the ``C" shape. (d)-(e) Different bending configurations of the "S" shape. (f) A non-planar ``S" shape configuration with corresponding displacement and force metrics. The vertical red dashed line represents the peak displacement, and the vertical green dashed line represents zero $\Delta F$.  }
\label{fig:penetration_test}
\end{figure*}

This experiment quantitatively evaluates the needle puncture performance and stiffness modulation capability of the proposed d-INA. The experimental setup, illustrated in Figure \ref{fig:penetration_test}(a), was designed to characterize the interaction between the d-INA and swine cardiac tissue by measuring the contact force and the catheter tip displacement before and after needle insertion.
The system consists of the outer sheath and the inner catheter assembled coaxially and passed through a plastic tube serving as a guiding channel that provides a well-defined boundary condition, allowing accurate assessment of the mechanical response of the bending segments. The entire assembly was mounted on an aluminum rail using 3D-printed brackets. Both the outer sheath and inner catheter can be independently rotated and translated within the guiding channel to achieve desired configurations of the bending segments. Once positioned, they can be secured using a screw-based locking mechanism to ensure stability during testing. To quantify the contact force between the catheter tip and the tissue, the swine cardiac tissue sample was mounted on a 3D-printed platform affixed to a high-resolution force sensor (Nano17, ATI Industrial Automation, USA). The combined platform and sensor assembly was secured to the aluminum rail, and for each tested configuration, the position and orientation of the force sensor platform were adjusted to ensure that the needle insertion direction remained aligned with the z-axis of the force sensor. To measure the catheter tip displacement, an electromagnetic (EM) tracking system (Aurora, NDI, Canada) was employed, with a tracking coil attached to the catheter tip to obtain its spatial position. 

Experiments were performed under five distinct configurations of the d-INA, including two ``C" shapes, two planer ``S" shapes, and one non-planar shape, as illustrated in Figure \ref{fig:penetration_test}(b)–(f). Prior to the needle insertion, the catheter tip was brought into gentle contact with the tissue, and the initial contact force (preload force) and catheter tip position were recorded. The needle was then manually advanced for 5 mm (maximum insertion length), and the updated force and position data were acquired. When the preload force was insufficient, the needle could not fully penetrate into the tissue, and the catheter tip was thus pushed back, indicating low stiffness in the needle insertion direction. 
The displacement of the catheter tip was quantified by comparing its position before and after each insertion. Following each attempt, the needle was fully retracted, completing one acquisition cycle. For subsequent cycles, the tissue sample was repositioned to ensure each puncture occurred at a fresh site. Then, fine adjustments were made to the tendon inputs of both the sheath and catheter to increase the preload force while maintaining the catheter tip position roughly unchanged. This is possible due to the fact that the device possesses 6 DoFs, which is redundant for maintaining 3-DoF tip position and thus can be exploited to change the preload force simultaneously, effectively modulating the stiffness along the needle direction. The insertion cycle was repeated as the preload force increased, until the needle could be fully inserted into the tissue.

% the preload force was increased by fine adjustments of the sheath and catheter handle, while maintaining the same bending orientation, and the insertion procedure and data acquisition were repeated. Such cycles were repeated with increasing preload force until the needle was able to be fully inserted into the tissue. 

Figure \ref{fig:penetration_test}(b)–(f) presents the tip displacement, contact force after needle insertion, and the corresponding difference between the preload force and the force after insertion ($\Delta F$), over various preload conditions for each configuration. 
In all cases, the displacement first increases up to a peak value as the preload force increases, after which it gradually decreases. %The force after insertion generally increases with the preload force, in most cases with a local peak at the displacement peak. \yc{why}
After the displacement peak, $\Delta F$ decreases asymptotically towards the horizontal zero line and intersects it. 
The observed trajectories delineate the sequential stages of needle penetration: prior to the displacement peak, the needle remains in contact with the tissue surface without penetration; between the displacement peak and the point where $\Delta F$ crosses zero, the needle is partially inserted into the tissue; and beyond that point, full insertion is achieved. These results indicate that the stiffness of the proposed d-INA device can be modulated along the penetration direction across multiple configurations, enabling reliable needle puncture performance.  

\subsection{Ex-Vivo Ablation Validation}

\begin{figure}[!t]
\centering
\includegraphics[width=\linewidth]{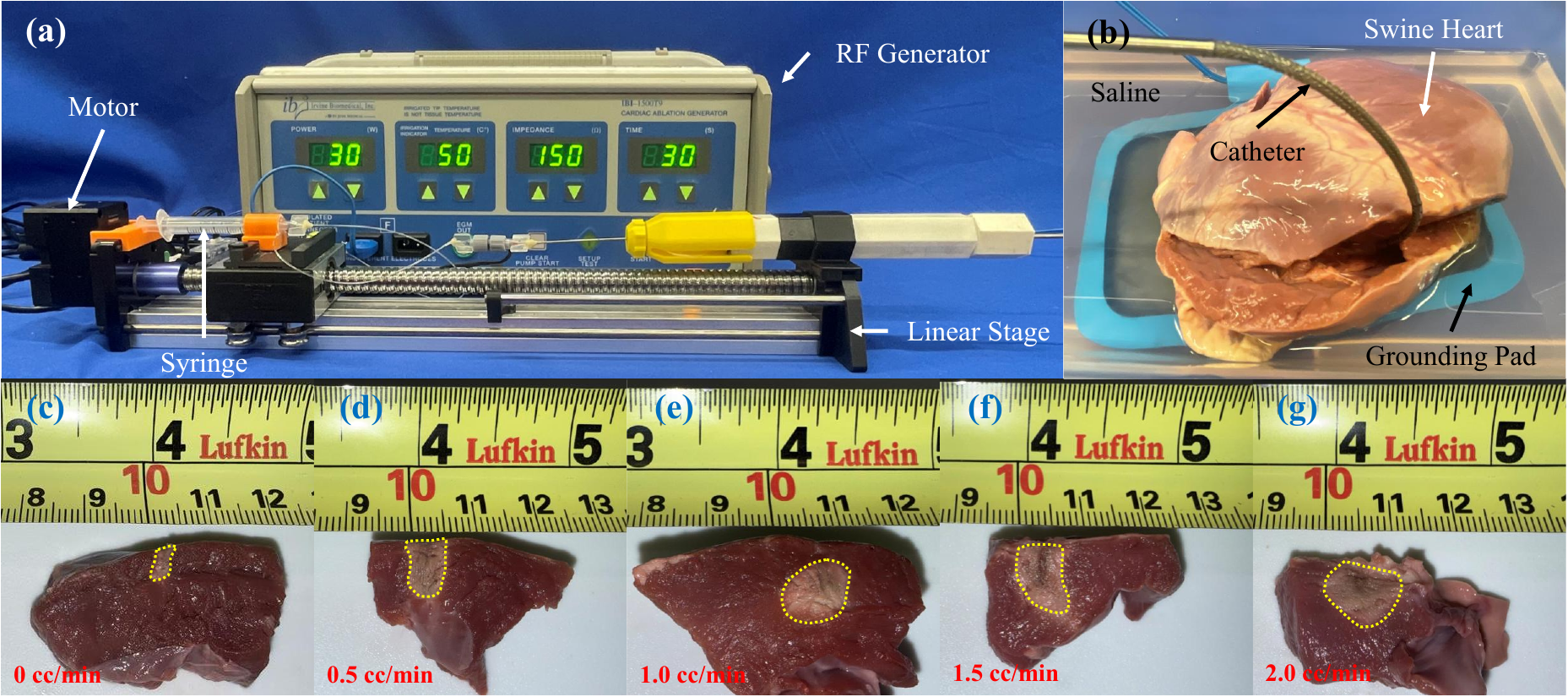}
\caption{The experimental setup and results of the irrigation varying ex-vivo ablation experiment. (a) The proximal end experimental setup. (b) The distal end experimental setup. (c)-(g) Ablation results using our proposed device under the ablation power of 30 W, duration of 30 s, and irrigation flow rate from 0 to 2.0 cc/min. The needle was inserted from top down.}
\label{fig:irrigation-ablation}
\end{figure}

To evaluate the ablation functionality of the proposed toolset, an ex-vivo experimental setup was developed, as illustrated in Figure \ref{fig:irrigation-ablation}(a). The barrel of a 3-cc syringe was fixed to the moving carriage of a linear motion stage and the plunger was secured to a stationary end of the stage, while the catheter handle was fixed to the other end. The catheter’s irrigation tube was connected directly to the syringe Luer hub. The motor actuation of the linear stage generated a controlled irrigation flow rate determined by the motor speed. The ablation wire of the needle was connected to a radio-frequency (RF) generator (IBI-1500T9, Irvine Biomedical Inc., USA). As shown in Figure \ref{fig:irrigation-ablation}(b), a grounding pad was submerged in saline together with the ex-vivo cardiac tissue, thereby completing the electrical circuit for ablation. The tissue ablation depth and area are dependent on three factors: power, time, and irrigation flow rate. Therefore, two experiments were conducted to assess these factors, including an irrigation flow rate test and a power-time evaluation. 

\subsubsection{Effects of Irrigation Flow Rate}

The irrigation flow rate was varied from 0 cc/min to 2.0 cc/min in the increment of 0.5 cc/min by changing the linear stage motor speed. At every irrigation flow rate, the needle was first inserted into the swine cardiac tissue from the surface with the insertion depth of 5 mm, and pre-hydration of the target tissue was conducted for one minute prior to the ablation. In all cases, the ablation power was 30 W and the duration was 30 s.  
After the ablation, the tissue was cut along the needle insertion line to evaluate the cross-section geometry of the ablation lesion, as shown in Figure \ref{fig:irrigation-ablation}(c)-(g). For the 0 cc/min case (Figure \ref{fig:irrigation-ablation}(c)), the absence of irrigation caused an impedance surge, which triggered a failsafe in the generator and terminated the ablation. Consequently, the 0 cc/min setup was only able to ablate for the first 6 s of the planned 30 s. Measurements of lesion width and height, presented in Figure \ref{fig:irrigation-ablation}, indicate that increasing the irrigation rate results in larger ablation volumes. At an irrigation flow rate of 2.0 cc/min, the proposed d-INA toolset produced lesions 10.4 mm deep in the swine cardiac tissue. 

\begin{figure}[!t]
\centering
\includegraphics[width=\linewidth]{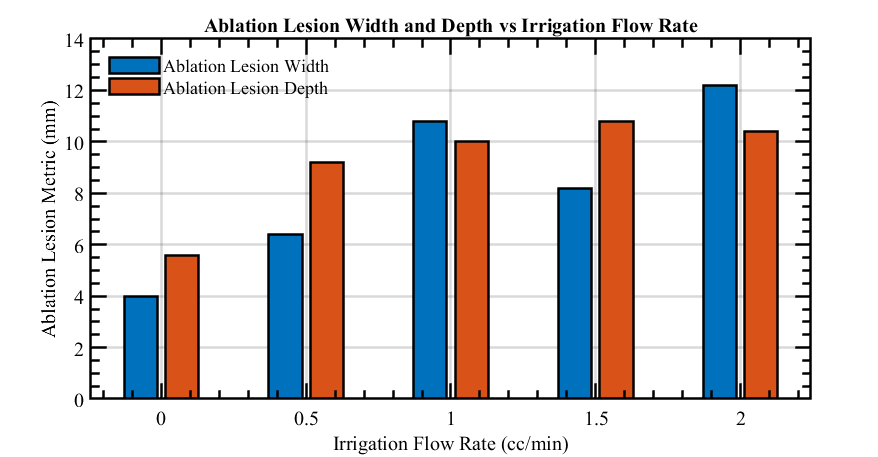}
\caption{The experimental results of the ablation lesion width and depth vs the irrigation flow rate. }
\label{fig:irri_ablation_results}
\end{figure}

\subsubsection{Effects of Ablation Energy} 

\begin{figure}[!t]
\centering
\includegraphics[width=\linewidth]{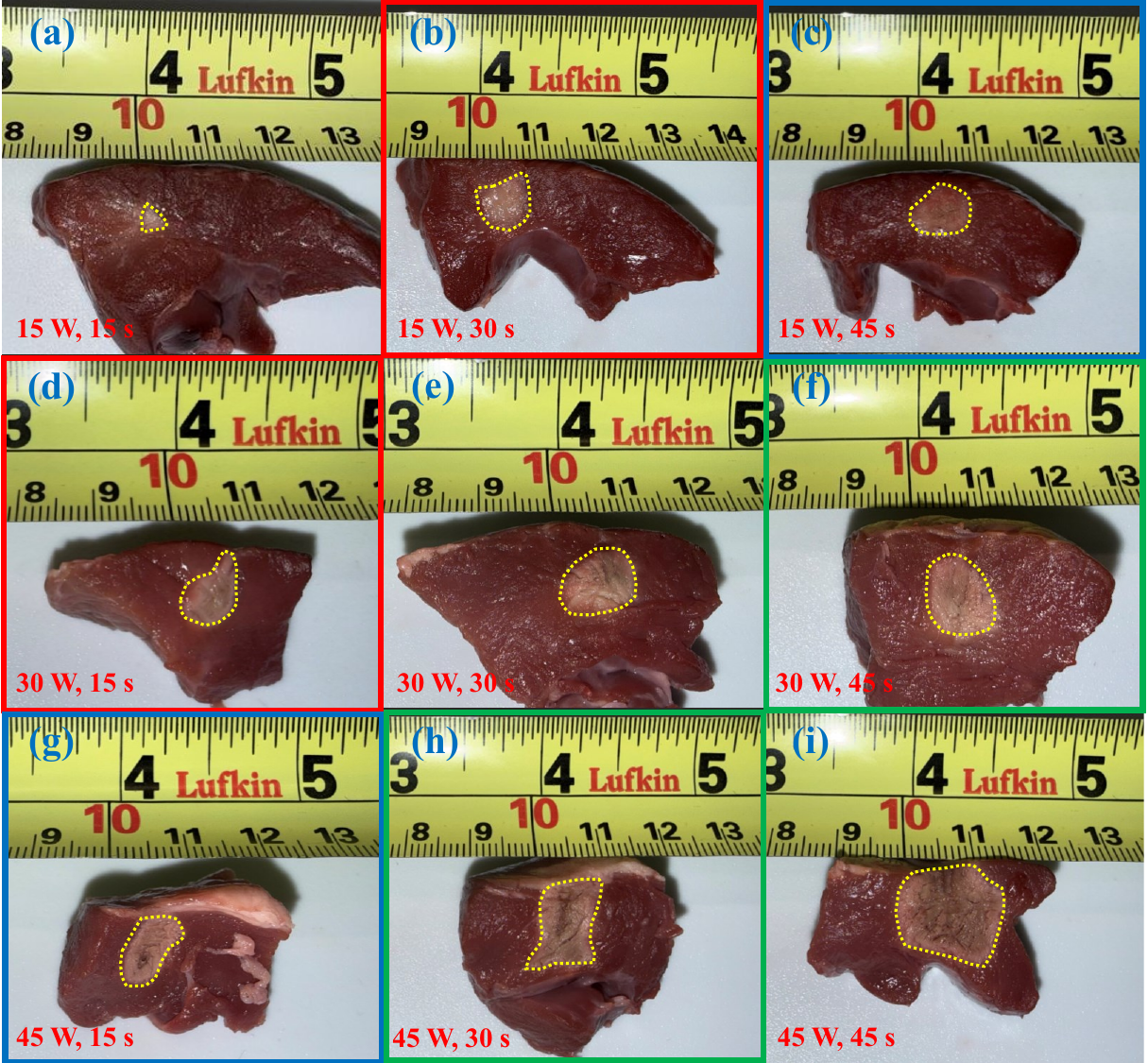}
\caption{The experimental results of the ablation power and time-varying ablation experiment. Test pairs with the same theoretical ablation energies are shown in bounding boxes of the same color (red, green, or blue). }
\label{fig:ablation_power_time}
\end{figure}

\begin{figure}[!t]
\centering
\includegraphics[width=\linewidth]{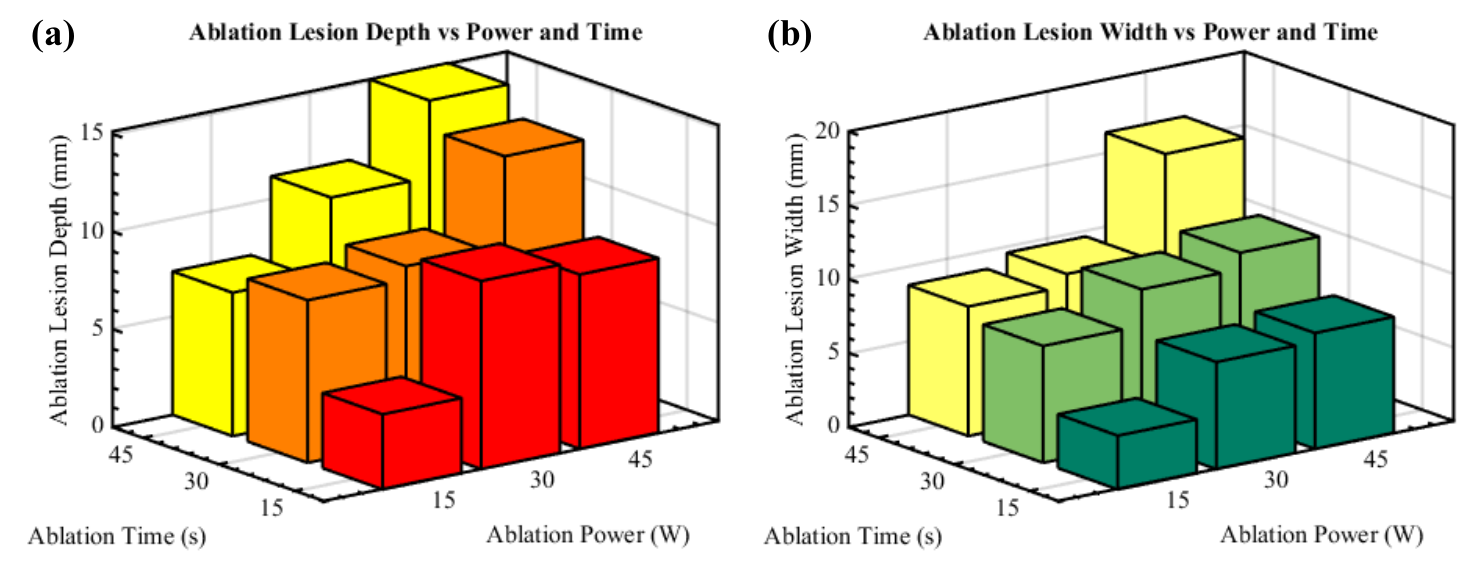}
\caption{The experimental results of the ablation lesion (a) depth and (b) width vs the ablation power and time.}
\label{fig:ablation_power_time_results}
\end{figure}

The energy delivered to ablate the targeted tissue depends on both power and ablation time. Both independent parameters were varied from 15 (watts or seconds) to 45 (watts or seconds) in increments of 15 (watts or seconds) per clinician’s prior experience. Irrigation rate was maintained at 1 cc/min for all experiments. The ablation energy study followed the same procedure described above. The resulting ablation lesions are shown in Figure \ref{fig:ablation_power_time}, and the corresponding lesion width and depth as functions of power and time are presented in Figure \ref{fig:ablation_power_time_results}. Increasing time and power led to an obvious increase in both the ablation lesion width and height. Notably, even for segments with equivalent total energy (e.g., Figure \ref{fig:ablation_power_time}(b) vs (d), (c) vs (g), (f) vs (h)), variations in the cross-sectional coloration were observed, with higher power levels producing greater thermal denaturation of the tissue, as evidenced by the dark charring core in the ablation lesions of Figure \ref{fig:ablation_power_time}(g)-(i).

\section{In-Vivo Experimental Validations}
\label{sec:InVivoExperimentalValidations}
To further evaluate the clinical viability of the d-INA system, swine studies were performed. All animal protocols were reviewed and approved by the Animal Care and Use Committee at the Johns Hopkins University (protocol number: SW23M128). The proposed d-INA performance was evaluated for performing RF ablation in naïve swine (20-30 kg weight). The procedures had diagnostic and interventional components and were conducted via X-ray fluoroscopy using a mobile C-ARM (GE OEC 7800, Salt Lake, UT) and an MRI scanner (Siemens 1.5T Aera). 

\subsection{Fluoroscopy-guided d-INA in the LV}

Two swine were studied under general anesthesia with mechanical ventilation. Percutaneous femoral vein access was performed prior to insertion of an 18Fr or 22Fr introducer sheath.  Using X-ray fluoroscopy guidance, transeptal puncture from the right to the left atrium was performed with a commercial deflectable sheath (Agilis, Abbott Inc.) and transeptal needle (BRK-1, Abbott Inc.) After left atrial access, the deflectable sheath was exchanged for the d-INA steerable outer sheath over a guidewire.  While orienting the sheath toward the mitral valve, the steerable d-INA catheter was inserted through the steerable  outer sheath into left ventricle (LV).  The steerable outer sheath was then advanced over the catheter to the level of the basal LV prior to inserting the new steerable needle into the steerable catheter. 

Joint navigation of all three components was used to reach target ablation locations.  d-INA RFA lesions were first targeted for the basal LV summit region just below the aortic valve prior to ablation of other sites.  The LV summit region can be challenging to reach from transseptal access through the mitral valve and required a high degree of torquability and deflection of the proposed deflectable sheath and catheter. Once the desired location was reached, the needle was deployed, penetrating the myocardium wall to a device-limited insertion depth of 5 mm. Iodine contrast was then flowed through the contrast channel in the inner catheter to confirm needle deployment into the myocardium wall. Following confirmation, a conventional INA protocol was followed: First, 2 cc of saline was injected into the tissue through the needle lumen. RF ablation was performed using a commercial RF Ablation generator (Abbott Ampere, St Paul, MI) during continuous 2 cc/min manual saline injection (15 W for 30 s prior to 30 W for 30 s).

\subsection{d-INA Lesion Assessment by MRI and Pathology}
Following euthanasia, hearts were removed, and the extent of ablation injury was visualized by ex-vivo MRI and gross pathology.  Non-contrast T1 weighted ex-vivo MRI parameters were 3D-GRE, isotropic resolution 0.5 mm$^3$, TR/TR/FA/Pixel BW/Averages = 25ms/4.76ms/40$^\circ$/200 Hz/10.

\subsection{d-INA  v.s. Conventional RFA Comparison}

\begin{figure}[!t]
\centering
\includegraphics[width=\linewidth]{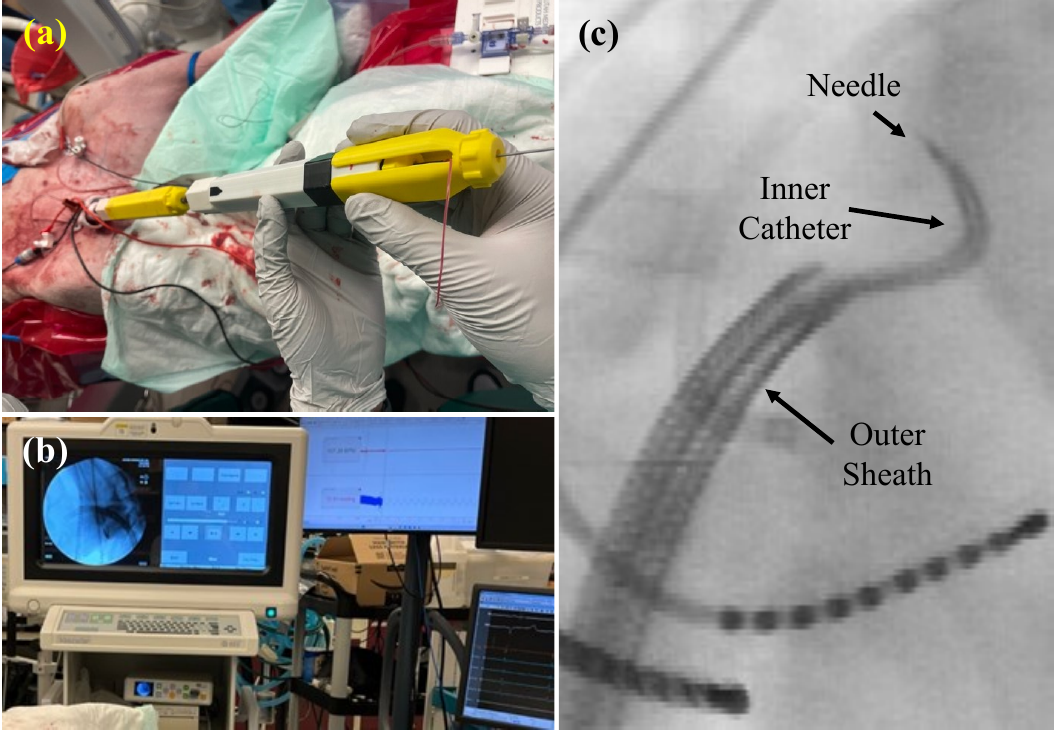}
\caption{Experiment setup in the fluoroscopy-guided INA in the left ventricle. (a) The insertion and navigation of the sheath and catheter. (b) Fluoroscopy image feedback of the toolset in-vivo. (c) The "S" shape configuration of the proposed d-INA toolset under fluoroscopic imaging.}
\label{fig:invivo_test}
\end{figure}

The d-INA prototype was evaluated under fluoroscopic guidance to explore three specifications of the system; (i) the ease of navigation with the proposed system, (ii) the ease of needle deployment at operator-selected targets of the LV, and (iii) the lesion geometry created by the INA delivery, relative to the conventional RF ablations, with equivalent RF generator power delivery.  

To study (i) the ease of navigation, the interventionalist navigated the d-INA system’s outer deflectable sheath together with its internal deflectable catheter through the mitral valve into the mid-LV and then made a sharp upwards curve with the protruding internal catheter, aiming to reach the basal anterior LV around the LV summit, which can be a difficult region to reach via transseptal access. Due to the ability to perform small radius deflections with the outer sheath, it was possible to bring the catheter to the two desired locations and with sufficiently perpendicular position to the LV wall to (ii) easily deploy the innermost needle into the myocardial wall. 

To validate the use of the d-INA system to create (iii) effective RF ablations in-vivo, the system was connected to the irrigation system and then to the RF generator, and ablation lesions were created. For comparison purposes, additional lesions were created with a conventional RF ablation catheter (Tacticath, Abbott Inc.) from the endocardial surface. In one animal, two INA and two conventional RFA lesions were created. In the second animal, four INA lesions and one conventional RFA lesion were created.   

\begin{figure*}[!t]
\centering
\includegraphics[width=0.75\linewidth]{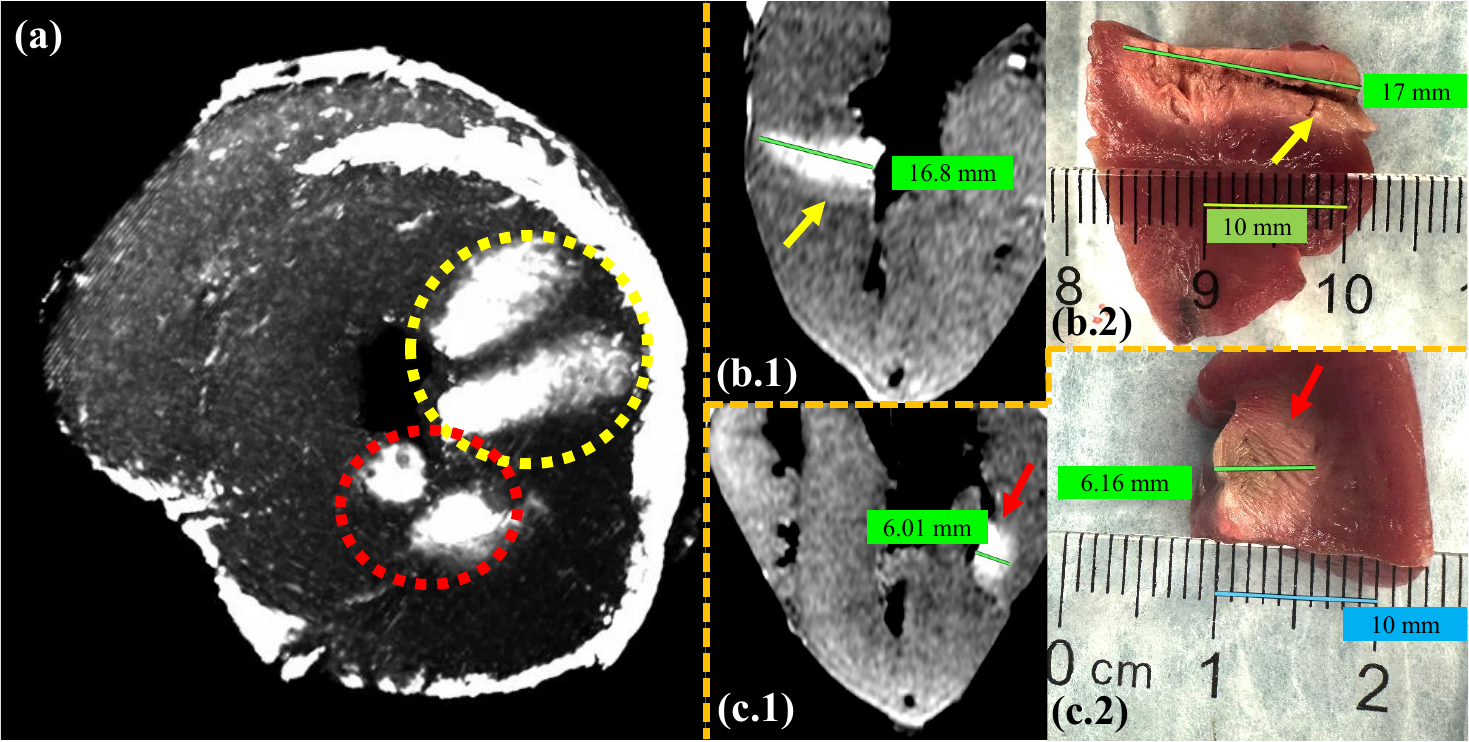}
\caption{MRI and pathology assessment of lesions created by prototype INA catheter vs. conventional RFA.  (a) MRI Volume rendering of ex-vivo non-contrast T1 weighted MRI shows lesion distribution after two INA lesions (yellow circle) and two conventional RFA lesions (red circle). Lesion depth from the endocardium measured by ex-vivo MRI, corresponds to pathology cross section, and is greater for INA (b.1), (b.2) than conventional ablation (c.1), (c.2) using same 30 W RF power for 60 s.}
\label{fig:invivo_result}
\end{figure*}

Figure \ref{fig:invivo_result} illustrates two d-INA ablation lesions compared with two endocardial-surface-based RFA lesions imaged by T1-weighted ex-vivo MRI (Figure \ref{fig:invivo_result}(a), (b.1), and (c.1)) and comparison to gross pathology lesion sections (Figure \ref{fig:invivo_result} (b.2) and (c.2)). 
%
%Comparatively, d-INA achieved a 282\% increase in ablation depth compared to the conventional RFA lesions. All INA lesions achieved more than 80\% lesion transmurality from endocardium to epicarduam, while no RFA lesion was more than 60\% transmural.
Across both swine, ablation lesion depth was 11.6+/-3mm for the d-INA vs. 5.6+/-0.4mm for conventional RFA.  Lesion transmurality from endocardium to epicardium was 92\% +/- 11\% for the d-INA vs. 42\%+/-6\% for conventional INA.  Comparatively, d-INA achieved a 219\% increase in ablation depth compared to the conventional RFA lesions.

%This result supports prior canine results using a commercial INA catheter \cite{dickow2020characterization}, demonstrating that the d-INA catheter can produce equivalent lesions to industrially built INA catheters. That study showed that if equivalent contact force was applied when the lesion was created, the mean depth of the INA ablated lesions reached 90\% of the LV wall in both normal and infarcted myocardium, and that INA lesions were transmural in 59.1\% versus 7.7\% in endocardial-surface-based RF ablation. 

This result supports prior canine results using a commercial INA catheter \cite{dickow2020characterization}, demonstrating that the d-INA catheter can produce equivalent lesions to industrially built INA catheters. That study showed that if equivalent contact force was applied when the lesion was created, the mean depth of the INA ablated lesions reached 90\% of the LV wall in both normal and infarcted myocardium, and that INA lesions were transmural in 59.1\% versus 7.7\% in endocardial-surface-based RF ablation.

% \section{Discussion}
% \label{sec:Discussion}
% \input{section/4.discussion.tex}

\section{Conclusions and Future Work}
\label{sec:conclusion}
This paper presented a novel d-INA device which is capable of dexterous motion and navigation in-vivo for deep-seated LV ablation. Our work addresses the critical need for greater steerability capable of reaching distant sections in the heart, increased torquability to control through the confined workspace of the heart, as well as stiffness modulation of the device for needle tissue penetration in any configuration. The d-INA consists of a steerable sheath, a steerable catheter, and an ablation needle. The combination of the sheath and catheter achieves both planar and non-planar shapes due to the proposed passive bending segment and steerable segment in the catheter. 
Both steerable devices showed exceptional curvature (0.088 $\text{mm}^{-1}$ for the sheath and 0.144 $\text{mm}^{-1}$ for the catheter). Additionally, the toolset exhibited the capability to achieve tissue penetration across multiple configurations through stiffness modulation. The toolset ablation was first verified in a benchtop ex-vivo tissue experiment. 
After acquiring satisfying results, the system was validated in an animal trial, where multiple lesions were created and assessed by ex-vivo MRI and pathology. The d-INA toolset achieved 219\% greater ablation depth compared to the conventional catheter counterpart. While the device has only been validated in limited in-vivo trials, it demonstrated the potential to reach challenging locations and creating deep-seated, transmural ablation lesions, addressing important limitations in the current VT therapy. Overall, we demonstrated the potential of the d-INA for the treatment of VT. 

Despite successful benchtop and in-vivo trials, several limitations were identified to guide future development.
First, the  workspace analysis based on the constant curvature assumption may prove insufficient in the practical environment. 
This is because  anatomical obstacles such as papillary muscles, mitral valve, and trabeculations can significantly alter the d-INA's shape and reachable workspace. To address this, future work will focus on developing mechanics-based models \cite{faure2012sofa} that account for these environmental interactions. Coupling this mechanics model with an optimal design process will yield a system that is more clinically viable. 
Second, manually operating the d-INA within  LV is challenging due to limited real-time feedback (see Fig.\ref{fig:invivo_test}a). Since TWILITE MRI has demonstrated efficacy in assessing ablation lesion quality \cite{guttman2020acute}, we envision a more comprehensive, MRI-based, robot-assisted closed-loop solution. A future clinical workflow could integrate pre-operative MRI planning, intraoperative real-time device tracking and robotic control \cite{chen2019closed, alipour2019mri}, multi-modal feedback such as temperature, force, and shape \cite{chen2014mri}, and immediate post-operative ablation assessment via TWILITE. This integrated system holds the potential to enable precise manipulation of the d-INA to target sites and the delivery of sufficient RF energy to create transmural necrosis, forming a robust platform for closed-loop VT ablation.

\bibliographystyle{IEEEtran}
\bibliography{root}

\end{document}